\documentclass{aa}  
\usepackage{graphicx,natbib}
\usepackage{txfonts}
\usepackage{psfig}
\newcommand{\Ha} {H$\alpha$}  
\begin{document}
\title{Is \object{BL Lacertae} an ``orphan'' AGN?}

\subtitle{Multiband and spectroscopic constraints on the parent
  population\thanks{Based on observations made with the Italian Telescopio
    Nazionale Galileo operated on the island of La Palma by the Centro Galileo
    Galilei of INAF (Istituto Nazionale di Astrofisica) at the Spanish
    Observatorio del Roque del los Muchachos of the Instituto de Astrof´ısica
    de Canarias.}}

   \author{A.~Capetti      \inst{ 1}
   \and   C.~M.~Raiteri    \inst{ 1}
   \and   S.~Buttiglione   \inst{ 2,3}
 }

   \offprints{A.~Capetti; \email{capetti@oato.inaf.it}}

   \institute{
          INAF, Osservatorio Astronomico di Torino, Via Osservatorio 20, I-10025 Pino Torinese, Italy                                                     
   \and   SISSA-ISAS, Via Beirut 2-4, I-34014 Trieste, Italy  
\and INAF, Osservatorio Astronomico di Padova, Vicolo dell'Osservatorio 5,
I-35122 Padova, Italy}

   \date{}
 
  \abstract
   {}
   {We have analysed optical spectra of BL Lacertae, the prototype of its
     blazar subclass, to verify the broad H$\alpha$ emission
     line detected more than a decade ago and its possible flux variation. We
     used the spectroscopic information to investigate the question of the BL
     Lacertae parent population.}
   {Low- and high-resolution optical spectra of BL Lacertae were acquired with
     the DOLORES spectrograph at the 3.58 meter Telescopio Nazionale Galileo
     (TNG) during four nights in 2007--2008, when the source was in a
     relatively faint state.  In three cases we were able to fit the complex
     H$\alpha$ spectral range with multiple line components and to measure
     both the broad H$\alpha$ and several narrow emission line fluxes.}
   {A critical comparison with previous results suggests that the broad
     H$\alpha$ flux has increased by about $50\%$ in ten years.  This might be
     due to an addition of gas in the broad line region (BLR), or to a
     strengthening of the disc luminosity, but such flux changes are not
     unusual in Broad Lined active nuclei. We estimated the BL Lacertae black
     hole mass by means of its relation with the bulge luminosity, finding
     4--$6 \times 10^8 \, M_{\sun}$.  The virial mass estimated from the
     spectroscopic data gives instead a value 20--30 times lower. An
     analysis of the disc and BLR properties in different AGNs suggests that
     this discrepancy is due to an underluminosity of the BL Lacertae BLR.
     Finally, we addressed the problem of the BL Lacertae parent population,
     comparing its isotropic quantities with those of other AGN classes. From
     the point of view of the narrow emission line spectrum, the source is
     located close to low-excitation radio galaxies. When one also considers its
     diffuse radio power, an association with FR~I radio galaxies is
     severely questioned due to the lower radio luminosity (at a given line
     luminosity) of BL Lacertae. The narrow line and radio luminosities of BL
     Lacertae instead match those of a sample of miniature radio galaxies,
     which however do not show a BLR. Yet, if existing, ``misaligned BL
     Lacertae" objects should have entered that sample. We also rule out the
     possibility that they were excluded because of a QSO optical
     appearance.}
   {The observational constraints suggest that BL Lacertae is caught in a
     short term transient stage, which does not leave a detectable evolutionary
     ``trace'' in the AGN population. We present a scenario that can
     account for the observed properties.}


   \maketitle

\section{Introduction}

According to the commonly accepted scenario, the central engine of active
galactic nuclei (AGNs) is a supermassive black hole (SMBH) fed by infall of
matter from an accretion disc. Inner fast-moving clouds produce broad emission
lines, which may be obscured by absorbing material, while outer clouds are
responsible for narrow emission lines.  About one fifth of AGNs is radio-loud
\citep{kellermann94},
showing plasma jets sometimes extending on Mpc scales. Among them, BL Lac
objects and flat spectrum radio quasars (FSRQs) form the blazar class,
characterized by variable emission from the radio to the $\gamma$-ray band,
with flux variations on time scales from hours to years, high radio and
optical polarization, core-dominated radio morphology, flat radio spectra, and
apparent superluminal motion of radio jet components. Their properties are
explained in terms of plasma relativistic motion in a jet pointing at a small
angle with the line of sight, with consequent beaming of the observed
radiation \citep{bla78}. Hence, the continuum radiation of blazars is
dominated by the relativistically beamed non-thermal radiation from the jet.

In FSRQs, thermal emission from the disc may be observable in the optical--ultraviolet band when the source is not in a flaring state. Disc signatures were detected e.g.\ in 3C 273 \citep{smi93,von97,gra04,tur06}, 3C 279 \citep{pia99}, 
PKS 1510-089 \citep{kat08,dam09}, 3C 345 \citep{bre86}, and 3C 454.3 \citep{rai07b,rai08c}.
Moreover, strong broad and narrow emission lines are usually present in their spectra.

As for BL Lac objects, according to the original definition, they may show at
most weak emission lines, with equivalent widths not exceeding 5 \AA\ in the
rest frame \citep{sti91}.  This seems to be due not so much to low line
fluxes, but rather to a high continuum flux \citep{sca97}.  Indeed, strong
emission lines, in particular broad ones, have occasionally been detected in
the spectra of BL Lac objects in faint states.  These include BL Lacertae, the
prototype of the blazar subclass named after it \citep{ver95,cor96,cor00},
and the distant source AO 0235+164 \citep{coh87,nil96,rai07a}.

The unified scheme for radio-loud AGNs predicts that BL Lac objects and FSRQs
are the beamed counterparts of Fanaroff-Riley type I (FR I) and Fanaroff-Riley
type II (FR II) radio galaxies, respectively, even if these correspondences were questioned by various observing evidences
\citep[e.g.][]{tad08}. In particular, many BL Lac
objects show high, FR II-like extended radio powers and morphologies
(see e.g. \citealt{lan08} \, and references therein).
The distinction between
blazars and radio galaxies, as well as that between FSRQs and BL Lac objects,
has been discussed by several authors, and different criteria have been
proposed, involving the value of the Ca H\&K break \citep{mar96}, or the
strength of the oxygen-narrow emission-lines \citep{lan04}.

A powerful way to classify AGNs is their position in diagnostic diagrams
comparing selected emission line ratios \citep{hec80,bal81}. In particular,
ratios of lines close in wavelength, like [\ion{O}{III}]/H$\beta$,
[\ion{N}{II}]$\lambda 6583$/H$\alpha$, [\ion{S}{II}]$\lambda \lambda 6716,
6731$/H$\alpha$, and [\ion{O}{I}]/H$\alpha$ are expected to be the most
reliable ones \citep{vei87}.  The application of diagnostic diagrams to
radio-loud galaxies by \citet{lai94} confirmed former suggestions that FR II
sources can be divided between high-excitation galaxies (HEG) and
low-excitation galaxies (LEG).  In particular, \citet{but10} verified this
dichotomy when analysing the radio sources belonging to the well known 3CR
catalogue. They found prominent broad lines in a sub-sample of HEG, but not in
LEG. Moreover, they saw that HEG are associated with very powerful FR II only,
while LEG are spread on a wide range of radio powers, and can be of both FR II
and FR I type. Actually, the situation is even more complex, as the existence
of miniature radio galaxies, characterized by extremely low radio power,
relatively luminous narrow emission lines, and no BLR, demonstrates
\citep{bal09}.

An analysis of the spectroscopic properties of blazars to understand their
relationship with the radio galaxies (and other AGN classes) is not an easy
task, because the dramatic variability of the non-thermal continuum flux
strongly affects the appearance of lines, especially in BL Lac
objects. However, this analysis can help clarify whether blazars differ from
radio galaxies only for their orientation with respect to the line of
sight, or if they are intrinsically different sources.

We present spectroscopic observations of BL Lacertae carried out
in 2007--2008 with the 3.58 m Telescopio Nazionale Galileo (TNG) on the Canary
Islands, to address the problem of its parent population.  In the same period
BL Lacertae was the target of a multiwavelength campaign by the Whole Earth
Blazar Telescope\footnote{{\tt http://www.oato.inaf/it/blazars/webt/}} (WEBT),
also involving three pointings by the XMM-Newton satellite. The results of the
WEBT campaign have been presented by \citet{rai09}. The source was observed in
a relatively faint state at all wavelengths, and a UV excess was clearly
visible in the source spectral energy distribution (SED), which was
interpreted as the signature of thermal radiation from the accretion
disc. 

\section{Spectroscopic observations and data reduction}

   \begin{figure}
   \resizebox{\hsize}{!}{\includegraphics{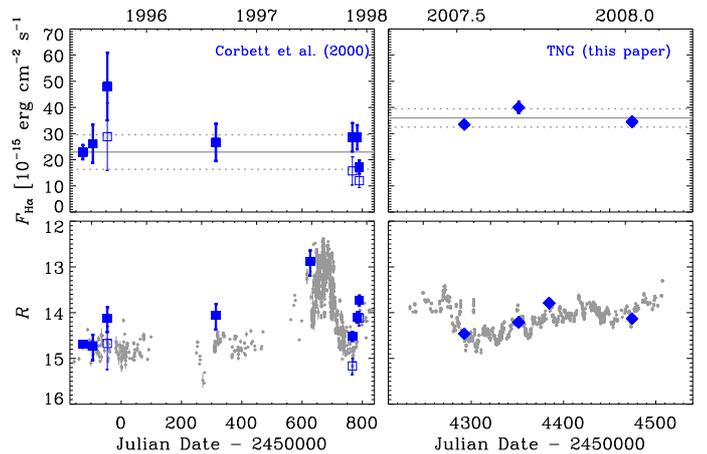}}
      \caption{The behaviour of the H$\alpha$ broad emission line flux as a 
function of time (top panels) compared to the continuum flux evolution (bottom panels, $R$-band magnitudes) in 1995--1997 (left) and 2007-2008 (right). Blue filled squares display the results of \citet{cor00}; blue empty squares represent data with revised flux calibration; blue diamonds show the results obtained in this paper by analysing the TNG data.
Grey dots represent data from the WEBT collaboration \citep{vil04a,rai09}.
The solid horizontal lines in the upper panels indicate the average H$\alpha$ flux in the corresponding period, while dotted lines are drawn through one standard deviation from the mean value.} 
         \label{cl}
   \end{figure}

   All spectra were taken with the Telescopio Nazionale Galileo (TNG), a 3.58
   m optical/infrared telescope located on the Roque de los Muchachos in La
   Palma Canary Island (Spain).  The observations were made with the DOLORES
   (Device Optimized for the LOw RESolution) spectrograph installed at the
   Nasmyth B focus of the telescope. The detector was a $2048 \times 2048$
   pixels E2V 4240 back-illuminated CCD, with a pixel size of $13.5 \mu$m, and
   a scale of $0.252 \arcsec$ pixel$^{-1}$, which implies a field of view of
   $8.6 \arcmin \times 8.6 \arcmin$.
 
   The spectroscopic observations were performed in service mode on 2007 July
   10--11, September 7, and October 10, and on 2008 January 8, during a
   multiwavelength campaign by the WEBT; in particular, the first and last
   observations were contemporaneous to XMM-Newton pointings at the source
   \citep{rai09}.  Figure \ref{cl} (bottom right) shows the $R$-band light
   curve obtained by the 30 optical telescopes of the WEBT: the source showed
   a noticeable variability, with short-term (intra-day or inter-day)
   flickering superposed on a longer-term trend. Blue diamonds display the
   results of aperture photometry performed on TNG images acquired with the
   Cousins' $R$ filter just before/after the spectra, with the same
   prescriptions used to derive the WEBT magnitudes, i.e.\ an aperture radius
   of $8 \arcsec$, and background derived from a surrounding annulus with $10
   \arcsec$ and $16 \arcsec$ radii.  In this way, \citet{vil02} estimated that
   the measured flux density includes 60\% of the host galaxy flux. Table
   \ref{pho} reports the $R$-band magnitudes\footnote{Throughout the paper
       we report uncertainties at 1$\sigma$ confidence level.} derived from
   the TNG imaging frames according to the above WEBT prescriptions.  Figure
   \ref{cl} shows that the photometry on the TNG images agrees excellently with the WEBT data and helps to put the spectroscopic observations into context.

   \begin{figure*}[htbp]
    \centerline{
    \psfig{figure=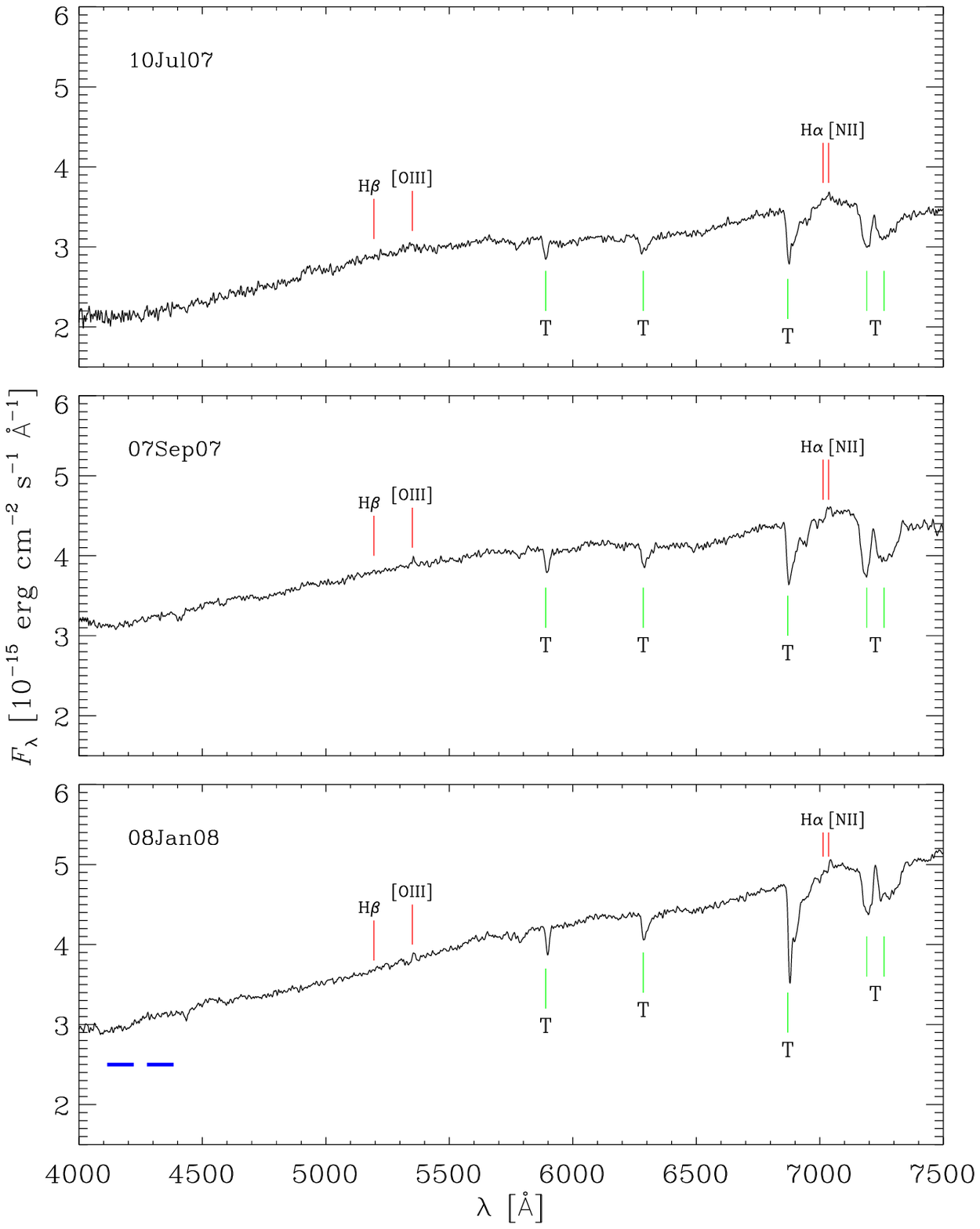,width=0.5\linewidth}
    \psfig{figure=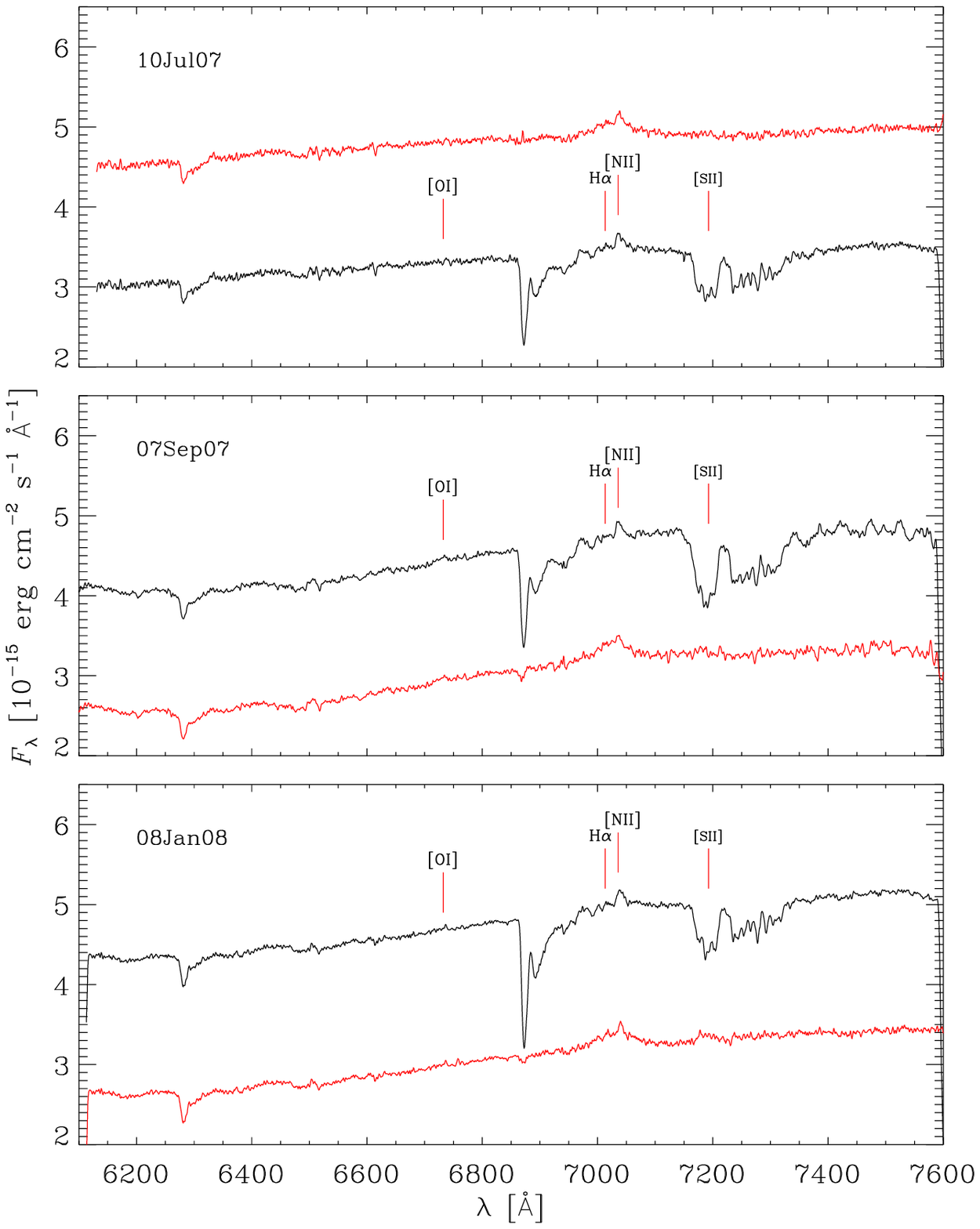,width=0.5\linewidth}}
  \caption{\label{spectra} {\it Left panels}: low-resolution TNG spectra.  The
    position of the emission lines is indicated as well as that of the
    telluric absorption lines.  Blue horizontal lines in the bottom panel mark
    the wavelength intervals used to calculate the D$_{n}$(4000) spectral
    index defining the strength of the Ca H\&K break.  {\it Right panels}:
    high-resolution TNG spectra both before (black) and after (red) correction
    for the telluric absorption lines around the H$\alpha$ line. The corrected
    spectra are shifted in flux for sake of clarity.}
   \end{figure*}

   These were performed with a $2 \arcsec$-width long-slit, which was aligned
   along the parallactic angle to minimize light losses due to
   atmospheric dispersion.  For each observing epoch, we obtained both
   low-resolution spectra with the LR-B grism, covering the wavelength range
   3500--7700 \AA\ with dispersion 2.52 \AA\ pixels$^{-1}$ and resolution 1200
   (for a $2 \arcsec$ slit), and high-resolution spectra with the VHR-R grism,
   sensitive to wavelengths from 6100 to 7700 \AA\ with dispersion 0.70 \AA\
   pixels$^{-1}$ and resolution 5000.  For each grism, two consecutive frames
   were taken and subsequently combined, for a total exposure of 600 s in
   LR-B, and 1400 s in VHR-R. The resulting signal-to-noise ratios of the
     co-added spectra are $\sim $ 50--100 and $\sim $ 100--200 for the low- and
     high-resolution spectra respectively.

   The use of both LR-B and VHR-R grisms allowed us to cover the spectral
   range where the most relevant emission lines of the optical spectrum are
   expected, in particular the key diagnostic lines H$\beta$,
   [O~III]$\lambda\lambda$4959,5007, [O~I]$\lambda\lambda$6300,64, H$\alpha$,
   [N~II]$\lambda\lambda$6548,84, and [S~II]$\lambda\lambda$6716,31.  The high-resolution spectra were needed to disentangle H$\alpha$ from the [N~II]
   doublet.

   The {\tt longslit} package of the NOAO's Image Reduction and Analysis
   Facility (IRAF)\footnote{IRAF is distributed by the National Optical
     Astronomy Observatories, which are operated by the Association of
     Universities for Research in Astronomy, Inc., under cooperative agreement
     with the National Science Foundation.}  was used to perform
   bias-subtraction, flat-field correction, wavelength calibration, background
   subtraction, spectra extraction (performed over 2$\arcsec$ in the spatial
   direction), and flux calibration.  For the last task we used the
   spectrophotometric standard star BD $+28\degr$4211 \citep{oke90}. We
   discarded the October spectra because of their low quality.

   \begin{figure}
   \resizebox{\hsize}{!}{\includegraphics{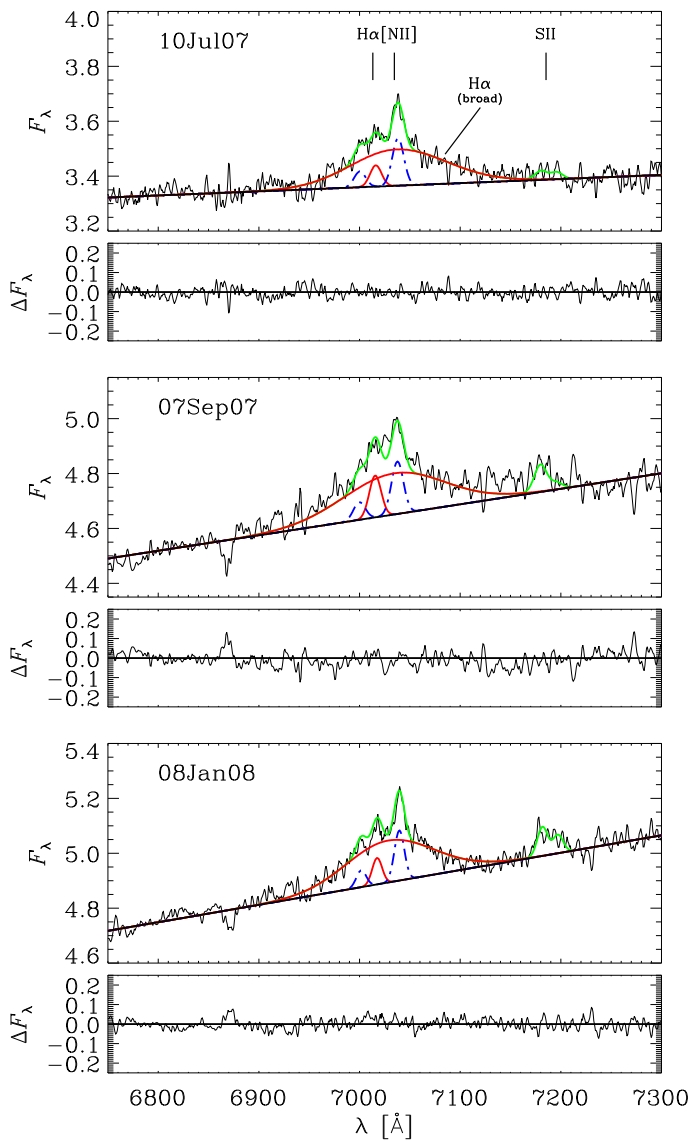}}
   \caption{Top panels: enlargement of the high-resolution, telluric-corrected
     spectra of Fig.\ \ref{spectra}, showing the results of the spectral
     fitting: red continuous lines indicate the fit to the broad and narrow
     components of the H$\alpha$ line; blue dotted-dashed lines show the fit
     to the [\ion{N}{II}] doublet; the green line displays the overall fit,
     which includes a tentative fit to the [\ion{S}{II}] line.  Bottom panels:
     residuals of the spectral fit.  Fluxes and residuals are given in
     $10^{-15} \rm \, erg \, cm^{-2} \, s^{-1} \, \AA^{-1}$ units.}
         \label{specfit}
   \end{figure}

   The absolute flux calibration of the spectra was checked through synthetic
   aperture photometry on the images taken in the Cousins' $R$ band just
   before/after the spectra. By setting an aperture size of $2 \arcsec$, we
   derived the flux densities at $\lambda_{eff} = 6410 \, \AA$ reported in
   Table \ref{pho}. The comparison with the spectra continuum around 6400 \AA\
   revealed that all spectra flux densities were in fair agreement with the
   fluxes inferred from photometry, with the only exception of the VHR-R
   spectra of July 2007. Indeed, the scaling factors to apply to the LR-B
   spectra were: 0.96, 0.95, 1.03 for July 2007, September 2007, and January
   2008, respectively, while for the VHR-R spectra we obtained: 1.27, 0.99,
   and 1.02.  For the LR-B spectra, which extend over almost all the
   wavelength range covered by the broad-band $R$ filter, these calibration
   factors were checked by convolving the BL Lacertae spectra with the
   transmission curve of the TNG $R$-band filter, multiplied by the CCD
   quantum efficiency curve.  We verified that the uncertainty on the
   calibration factors is of the order of 2\%.

\begin{table}
\caption{Results of aperture photometry on the TNG imaging frames.}
\label{pho}
\centering
\begin{tabular}{l c c }
\hline\hline
Date               & $R^{\rm a}$ & $F_R^{\rm b}$    \\
                   & [mag]       & [$10^{-15} \rm \, erg \, cm^{-2} \, s^{-1} \, \AA^{-1}$] \\
\hline
2007 Jul 10--11    & $14.46 \pm 0.01$ & $3.16 \pm 0.01$ \\
2007 Sep 7         & $14.21 \pm 0.02$ & $4.12 \pm 0.09$ \\
2007 Oct 10        & $13.79 \pm 0.01$ & $6.31 \pm 0.05$ \\
2008 Jan 8         & $14.13 \pm 0.03$ & $4.42 \pm 0.11$ \\
\hline
\end{tabular}

$^a$ $R$-band magnitudes obtained from photometry with $8 \arcsec$ aperture radius\\

$^b$ Flux densities at $6410 \, \AA$ derived from photometry with $2 \arcsec$ aperture radius, used to normalise the spectra.\\

\end{table}

By following \citet{sca00}, who gave the host galaxy brightness $R=15.55 \pm
0.02$ and its effective radius $r_e = 4.8 \arcsec$, and adopting a De
Vaucouleur brightness profile, we estimated that within $2 \arcsec$ the host
galaxy contribution is $\sim 0.32 \times 10^{-15} \rm \, erg \, cm^{-2} \,
s^{-1} \, \AA^{-1}$, i.e.\ $\sim 10\%$ of the observed flux in July, and $\sim
8\%$ and $\sim 7\%$ in September and January, respectively. 
This result is confirmed by the analysis of the Ca H\&K break
strength in the spectra. In the January spectrum we found that the ratio
between the fluxes in the rest-frame wavelength ranges 4000--4100 \AA\ and
3850--3950 \AA\ is $\rm D_{n}(4000) \sim 1.06$ \citep{bal99}.
Assuming $\rm D_{n}(4000) \sim 2$ for the host galaxy \citep{kau03}, as
typical of giant early-type galaxies, this
implies a host galaxy contribution not greater than a few percent.  Thus we
did not correct the spectra for the host galaxy contribution, because it is
negligible for our purposes.

The VHR-R spectra were corrected for the telluric absorption bands,
a particularly important step, because the 6870 \AA\ oxygen $B$-band affects
the blue wing of the H$\alpha$ line. This
was done by dividing the spectra by a template obtained from the
spectrophotometric standard star.

Figure \ref{spectra} displays the low-resolution (left panels) as well as the
high-resolution (right panels) spectra, the latter both before and after
correction for atmospheric absorption around the H$\alpha$ line. 

We measured the emission line intensities by means of the {\tt specfit}
package in IRAF \citep{kri94}.  To reduce the number of free parameters we
fixed the relative wavelength distance between lines and required the FWHM to
be the same for all the narrow lines. From the low-resolution spectra we
derived the [O~III]$\lambda\lambda$4959,5007 flux, and an upper limit to the
H$\beta$ flux.  In the high-resolution spectra, we fitted the 6700--7350 \AA\
spectral region with seven components: a Gaussian profile for each of the
lines: H$\alpha$ (broad and narrow), [N~II]$\lambda\lambda$6548,84, and
[S~II]$\lambda\lambda$6716,31, plus a linear component for the continuum.  The
results of this spectral fitting are shown in Fig.\ \ref{specfit}. The fit to
the [S~II]$\lambda\lambda$6716,6731 doublet must be considered with caution,
because this line is strongly affected by atmospheric absorption and thus its
measurement strongly depends on the telluric correction (see Fig.\
\ref{spectra}).
The FWHM of the narrow lines ranges from $\sim 500$ to $\sim 600 \rm \, km \,
s^{-1}$, while that of the broad H$\alpha$ component is 4200--5000 $\rm km \,
s^{-1}$.  Including a further component to fit the
[O~I]$\lambda\lambda$6300,64 line resulted in a marginal detection only.

Table \ref{lines} reports the line fluxes after correction for the Galactic
extinction, adopting $A_B = 1.42$, $E(B-V)$=0.35, and the extinction law of
\citet{car89}.

\begin{table*}
  \begin{center}
    \caption{Emission lines fluxes after correction for Galactic extinction. 
All fluxes are $10^{-15} \rm \, erg \, cm^{-2} \, s^{-1}$; FWHM are $\rm km \, s^{-1}$.}
    \label{lines}
    \begin{tabular}{l | c c c c c | c c}
      \hline \hline
Obs  & H$\beta$ & [O~III]$\lambda$4959,5007 & H$\alpha({\rm narrow})$  & [N~II]$\lambda$6548,84 &   [S~II]$\lambda\lambda$6716,6731& H$\alpha({\rm broad})$ & FWHM \\
\hline
2007 Jul 10--11 & $< 7.6$ & $4.7 \pm 1.2$ & $2.4 \pm 0.5$ & $5.3 \pm 0.6$ & $<$3.1        &  $33.5 \pm 0.9$ &  4600 \\
2007 Sep 7      & $< 3.0$ & $3.4 \pm 0.5$ & $4.8 \pm 1.5$ & $6.1 \pm 1.0$ & $3.9 \pm 0.9$ &  $40.0 \pm 2.1$ &  5000 \\
2008 Jan 8     & $< 2.0$ & $2.5 \pm 0.4$ & $2.5 \pm 0.6$ & $4.7 \pm 0.7$ & $4.4 \pm 0.8$ &  $34.5 \pm 0.9$ &  4200 \\
 \hline
    \end{tabular}
  \end{center}
\end{table*}

\section{Discussion}

\subsection{Variability of the broad H$\alpha$ emission line}

The first detection of broad emission lines in the optical spectrum of BL
Lacertae was reported by \citet{ver95}. They took two spectra in May and June
1995, and measured H$\alpha$ with FWHM of 3400--$3700 \rm \, km \, s^{-1}$ and
flux of $4.4 \times 10^{-14} \rm \, erg \, cm^{-2} \, s^{-1}$, while for H$\beta$
they obtained $ \rm FWHM = 4400 \, km \, s^{-1}$ and $F = 1.3 \times 10^{-14}
\rm \, erg \, cm^{-2} \, s^{-1}$. They stressed that such a luminous H$\alpha$
line should have been observed before, while it is not recognizable in earlier
spectra taken in 1975--1976 by \citet{mil77}, in 1985 by \citet{law96}, and in 1989 by \citet{sti93}. They estimated that the broad H$\alpha$ flux may have increased by at least a factor 5 since 1989.

Soon after, \citet{cor96} analysed two other optical spectra acquired in June
1995, and confirmed the results of \citet{ver95}. 
They discussed the appearance of the H$\alpha$ line
as due to an increase of either the amount of gas in the broad line region
(BLR) or the strength of the photoionising source. In the latter case, the
accretion disc seemed the most likely source of photoionising radiation. The
signature of the accretion disc was later recognized by \citet{rai09} as a UV
excess in the broad-band SEDs of BL Lacertae built with contemporaneous
low-energy data from the WEBT and high-energy data from the XMM-Newton
satellite.  

The spectroscopic monitoring of BL Lacertae was continued by
\citet{cor00}, with eight spectra taken between June 1995 and December 1997,
in seven of which they were able to measure the H$\alpha$ broad emission
line. Their results are reported in Fig.\ \ref{cl}, where the upper left panel
shows their line flux (corrected for extinction and host galaxy contamination)
as a function of time, while the lower left panel displays their continuum
estimates superposed to the light curve by the WEBT collaboration
\citep{vil04a}. They could not distinguish the H$\alpha$ line in June 1997,
when the source was experiencing a big optical outburst.  When considering
their estimates of the H$\alpha$ broad emission line flux together with ours,
we find an average flux of $30.6 \times 10^{-15} \rm \, erg \, cm^{-2} \,
s^{-1}$, and a standard deviation of $8.8 \times 10^{-15} \rm \, erg \,
cm^{-2} \, s^{-1}$.  This would imply that if we take into account the
errors, all H$\alpha$ fluxes are within one standard deviation from the mean
value, with the only exception of the December 1997 value.  Indeed,
\citet{cor00} concluded that in 1995--1997 there was no ``compelling evidence"
of ``any significant variation", and now we could add that after about 10
years the H$\alpha$ flux has still roughly the same intensity.

However, the WEBT photometry allows us to improve the \citet{cor00} absolute
flux calibration.  The spectra of August 24, 1995, November 14, 1997, and
December 7, 1997 are the ones for which we both have contemporaneous WEBT data
and found significant deviations of the \citet{cor00} measures from the WEBT
ones. The blue empty squares in Fig.\ \ref{cl} represent data rescaled; the
flux rescaling factors are 0.6, 0.55, and 0.7 for the three spectra,
respectively.  Thus, if we consider all the H$\alpha$ broad emission line
fluxes in the 1995--1997 period, rescaled to match photometric values,
we find an average value of $(23.0 \pm 6.6) \times 10^{-15} \rm \, erg \,
cm^{-2} \, s^{-1}$.  This can be compared to the average value of $(36.0 \pm
3.5) \times 10^{-15} \rm \, erg \, cm^{-2} \, s^{-1}$ obtained by considering
our three measurements in 2007--2008, suggesting that a $\sim 50\%$ increase
of the broad H$\alpha$ intensity may have occurred in a roughly 10 year time
interval. However, this is not unusual and it is not necessarily due to an
evolutionary trend. Indeed, oscillations of the H$\alpha$ flux up to $\sim
77\%$ on a time scale of a few years have been reported by \citet{kas00} for PG
quasars.

Due to its greater distance from the central engine, the narrow line region
(NLR) is expected to react to variations of the disc luminosity on much longer
timescales than those characterizing the BLR. Indeed \citet{cor96} reported
fluxes of the narrow H$\alpha$ line component and of [\ion{N}{II}] that are
inside the range of our results (see Table \ref{lines}). They also identified
[\ion{O}{I}] and [\ion{Fe}{VII}]. \citet{ver95} measured the narrow
[\ion{O}{III}] and [\ion{N}{II}] emission lines, with fluxes similar to ours.
[\ion{O}{III}] was previously measured also by \citet{sti93} in 1989, and by
\citet{law96} in 1985, who estimated a flux within a factor of 2 with respect
to our spectra. In conclusion the NLR luminosity did not undergo significant
changes in the last twenty years, although it must be noted that the narrow
line fluxes are known with larger uncertainties than the broad H$\alpha$.

\subsection{Black hole mass, BLR, and accretion disc properties}
\label{mbh}

The masses of SMBH powering AGNs can be estimated 
in a number of different ways.

One method relies on the presence of a relation between SMBH mass and
near-infrared bulge luminosity as derived by \citet{mar03}.  From the BL
Lacertae host galaxy brightness $R=15.55 \pm 0.02$ \citep{sca00}, correcting
for Galactic extinction and using a mean colour index for elliptical galaxies
of $R-K=2.71$ \citep{man01} we obtain an apparent magnitude $K=11.95$.  This
translates
into an absolute magnitude $M_K=-25.33$, and hence $\log
(L_K/L_{\sun,K})=11.44$. According to \citet{mar03}, this value implies
$M_{\rm BH} \sim 6 \times 10^{8} M_{\sun}$.

A revision of the SMBH mass versus bulge luminosity relation for AGNs has been
proposed by \citet{ben09}, using a sample of 26 objects observed by the Hubble
Space Telescope, and for which SMBH masses have been estimated through the
reverberation-mapping technique.  By adopting for BL Lacertae $L_V=5.55 \times
10^{10} L_{\sun}$ we derive $M_{\rm BH} = 3.76^{+1.28}_{-0.95} \times 10^{8}
M_{\sun}$, in substantial agreement with the previous estimate\footnote{The
  SMBH mass of BL Lacertae was also calculated by \citet{woo02} through the
  correlation between the SMBH mass and the stellar velocity dispersion
  $\sigma$; the estimate of $\sigma$ is inferred from the morphological
  parameters of the host galaxy, via the fundamental plane \citep{jor96}. They
  found $M_{\rm BH} = 1.70 \times 10^{8} M_{\sun}$.}.

Another possible approach is to use the spectroscopic information, and to
calculate the virial mass: $M_{\rm BH} = f R_{\rm BLR} \sigma_{\rm line}^2/G$,
where $f$ is a factor depending on the BLR structure, kinematics, and
orientation; $R_{\rm BLR}$ is the size of the BLR; $\sigma_{\rm line}$ is the
line dispersion, and $G$ is the gravitational constant.  Following
\citet{pet04}, $f=5.5$, and $\sigma_{\rm line}= {\rm FWHM}/2.355$.  Lacking
a measurement from reverberation mapping, the size of the BLR can be derived
from the scaling relationship with its luminosity as discussed by
\citet{kas05}. According to our data, the luminosity of the broad H$\alpha$
line is $\sim 4 \times 10^{41} \, \rm erg \, s^{-1}$; adopting a flux ratio
H$\alpha$/H$\beta\sim 3$ (an assumption supported by the measurements by
\citealt{ver95}), this leads to $R_{\rm BLR} \sim 5 \, \rm lt \, day$.  Taking
$\rm FWHM = 4600 \, km \, s^{-1}$ from Table \ref{lines}, we obtain $M_{\rm
  BH} \sim 2 \times 10^{7} M_{\sun}$. Following the prescription of
  \citet{marconi08} to account for the possible role of radiation pressure on
  the BLR, we derive a relatively small correction term to the black hole mass
  of $\sim 0.6 \times 10^{7} M_{\sun}$. The virial estimate of $M_{\rm BH}$
is then about 20--30 times lower than the values derived above.

At this stage the reason for the discrepancy is unclear.

Further clues on the properties of BL Lacertae come from the
relationship of the observed BLR with the accretion disc. 
The Optical Monitor XMM-Newton observations of BL Lacertae presented in
\citet{rai09} revealed a sharp up-turn in the SED
ultraviolet region, the characteristic signature of the Big Blue Bump associated with
the emission of the accretion disc. 
>From Figs\ 6 and 7 in \citet{rai09} one can estimate a disc
luminosity at 2500 \AA\ of $\log L_\nu (2500 \AA) \sim 29.4 \rm \, erg \, s^{-1} \, Hz^{-1}$. 
This value is similar to that measured in type 1 AGN (both radio-loud and radio-quiet) like Fairall 9, PG 1229+204, 3C 390.3, 3C 120, Mkn 509, PG 2130+099, PG
0844+349, and Akn 120 \citep{vas09}. The luminosity of the broad H$\beta$
lines of these sources ranges from 0.2 to $0.6 \times 10^{43} \rm \, erg \, s^{-1}$ \citep{kas05}.
For BL Lacertae, considering its broad H$\alpha$ line luminosity 
and the H$\alpha$/H$\beta$ flux ratio (see above), we obtain a factor of $\sim 15$--40 lower.

A comparison with broad-line radio galaxies in the 3CR sample \citep{but10}
gives a similar result: the ratio between the broad H$\alpha$ and the
rest-frame UV flux ranges between 100 and 250 \AA, while this ratio for BL
Lacertae is $\sim 0.3 \AA$. This difference is preserved considering an AGN of
very low BLR luminosity, $\sim 100$ smaller than BL Lacertae, the LINER galaxy
NGC 4579 \citep{bar96,bar01}, for which this value is $\sim 275 \AA$.
Apparently, the BLR of BL Lacertae is strongly underluminous with respect to
its disc emission when compared to other AGN.

This suggests a possible interpretation for the different values of SMBH mass
found above.  In fact, $M_{\rm BH}$ scales with the BLR luminosity as $M_{\rm
  BH} \propto L_{\rm BLR}^{0.6-0.7}$.  To account for an underestimate of the
SMBH mass by a factor of 20--30, the BLR should be underluminous by a factor of
70--300, in broad agreement to what we derived considering the ratio between
BLR and UV fluxes.

\subsection{The parent population}

We here explore how BL Lacertae would look like when seen with its jet
pointing at a larger angle from our line of sight, i.e. what extragalactic
sources might represent the misoriented parent population.
We thus consider all quantities that are not affected by beaming and that
might eventually depend on orientation only if there is selective
obscuration, as e.g.\ due to a flattened circumnuclear dust
structure, e.g.\ a torus.

The host of BL Lacertae is a giant elliptical galaxy of absolute magnitude
$M_K=-25.33$ (see Sect.\ref{mbh}), nearly two magnitudes brighter than the
characteristic absolute magnitude $M^{*}$ in this band \citep{sch76,hua03}.
The emission from the accretion disc, to which we associate the emission in
the near UV band at 2500 \AA, amounts to $10^{29.4} \rm \, erg \, s^{-1} \,
Hz^{-1}$.  In order to estimate the contrast between the disc and the host we
used the \citet{rai09} black-body fit to the accretion disc emission, which
gives an $R$-band flux of $0.95 \times 10^{-15} \rm \, erg \, cm^{-2} \,
s^{-1} \, \AA^{-1}$ The host-galaxy observed magnitude $R=15.55$
\citep{sca00}, after correcting for Galactic extinction, translates into a
flux of $3.07 \times 10^{-15} \rm \, erg \, cm^{-2} \, s^{-1} \, \AA^{-1}$.
This implies a ratio between host and nucleus of $\sim 3$, similar to that
measured in Seyfert 1 and Broad Line radio galaxies \citep{bentz09b}.

The extended radio emission is also unaffected by orientation.  By means of
VLA maps at 20 cm, \citet{ant85} estimated an extended radio flux of $40 \,
\rm mJy$. More recent VLA observations at 20 cm for the MOJAVE project
\citep{coo07} resulted in an extended flux of $18 \, \rm mJy$, essentially
confirming the earlier results.  These high spatial resolution observations
might, in principle, have missed diffuse low-surface brightness,
steep-spectrum emission.  We then consider the 74 MHz flux density measured
for BL Lacertae in the VLA Low-frequency Sky Survey (VLSS, \citealt{coh07}),
i.e.\ $F_{74} = 1.46 \, \rm Jy$.  Even in the assumption that the radio core
does not contribute significantly at this low frequency, this translates into
an upper limit of $\sim 180 \, \rm mJy$ at 1400 MHz (having adopted a radio
spectral index of 0.7). This is an extremely conservative limit, considering
that its radio core has a typical flux of $\sim$ 2 Jy (see
e.g. \citealt{kharb10}). Thus we are confident that a value of 40 mJy at 1400
MHz is well representative of the total extended emission in BL Lacertae. This
corresponds to a radio luminosity power of $\log P_{\rm ext}= 30.57 \, \rm erg
\, s^{-1} \, Hz^{-1}$.

\begin{figure}
   \resizebox{\hsize}{!}{\includegraphics[angle=90]{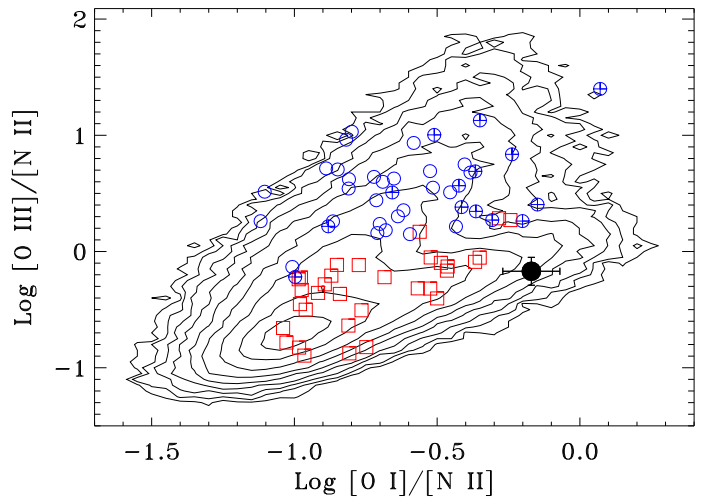}}
   \caption{The diagnostic index $\log \rm [\ion{O}{III}]/[\ion{N}{II}]$
     vs. $\log \rm [\ion{O}{I}]/[\ion{N}{II}]$ of BL Lacertae derived from our
     TNG data and those of \citet{cor96} (black dot). The
     red squares and blue circles show the distribution of LEG and HEG,
     respectively, according to the results by \citet{but10}. The crossed
     circles mark HEG that also show a BLR. Contours represent
   the density of SDSS emission line galaxies from \citet{kew06}.}
         \label{diag}
   \end{figure}

>From the point of view of the emission lines, we derived a broad H$\alpha$
luminosity of $\sim 4 \times 10^{41} \, \rm erg \, s^{-1}$ (see Sect.\ref{mbh}).
Considering the narrow emission lines, from Table \ref{lines} we estimate a
[O~III] luminosity of $\sim 4 \times 10^{40} \, \rm erg \, s^{-1}$.  The
narrow emission line ratios can contribute to characterize the properties of
an AGN, but unfortunately accurate measurements from our data are available
only for [O~III] and [N~II]\footnote{The narrow component of H$\alpha$ is
  heavily blended with the brighter broad emission; the [S~II] doublet is
  significantly uncertain as it falls deep into a telluric band; our data
  provide us with an upper limit on H$\beta$ and only with a tentative
  detection of [O~I]$\lambda$6300.}. As discussed in Sect.\ 3.1, the narrow
emission lines do not seem to have varied significantly in the last 20 years,
so we can complement our data with those from the literature. However, the
spectrum of \citet{ver95} does not improve the situation, because the H$\beta$
line is detected, but a decomposition into narrow and broad component could
not be performed.  The observations by \citet{cor96} cover only the red part
of the spectrum ($\lambda \gtrsim 6000 \AA$) where they saw a rather
well-defined [O~I]$\lambda$6300 line. This enables us to locate BL Lacertae in
a non-standard diagnostic plane defined by the ratios [O~I]/[N~II] and
[O~III]/[N~II] (Fig.\ \ref{diag}). As a comparison, we show in this diagram
the location of the 3CR sources (limited to a redshift of 0.3) from
\citet{but10}. The two main spectroscopic classes of Low and High Excitation
Galaxies (LEG and HEG, respectively) define two separate sequences, which are
also present in the SDSS emission line galaxies from \citet{kew06}. BL
Lacertae falls in a region not well populated by 3CR sources, but it is closer
to LEG than to HEG; furthermore it lies on the branch of the LINERs from the
SDSS. This suggests a tentative identification as a Low Excitation Galaxy
from the point of view of its narrow emission line spectrum.

\begin{figure}
   \resizebox{\hsize}{!}{\includegraphics[angle=90]{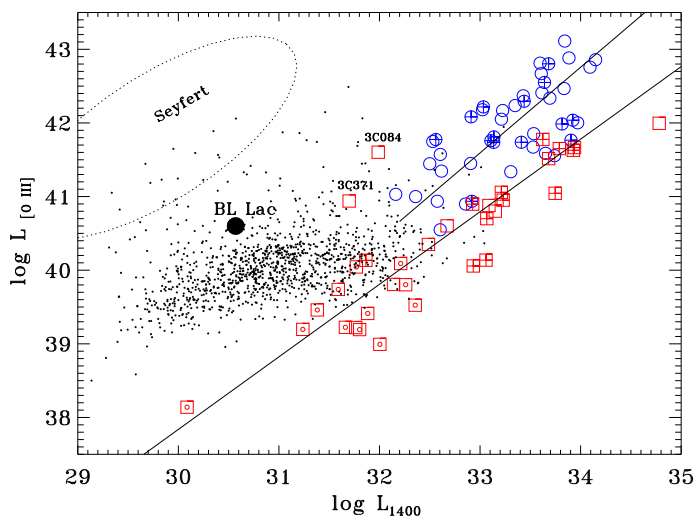}}
   \caption{ [O~III] luminosity [erg s$^{-1}$] as a function of radio
     luminosity at 1400 MHz [erg s$^{-1}$ Hz$^{-1}$]. The large black dot
     represents the location of BL Lacertae. Blue circles are HEG (crossed
     circles are broad line objects), while red squares are LEG.  
     When possible, we further
     mark LEG according to their FR type: crossed squares are FR~II/LEG
     and dotted squares are FR~I/LEG. The two solid lines represent the best
     linear fit obtained for the HEG and LEG sub-populations separately. The
     small black dots are the radio-sources from the SDSS/NVSS sample (see
     text for details). The dotted ellipse includes the
Seyfert galaxies considered by \citet{whittle85}.}
\label{3c}
   \end{figure}

   Which class of objects share the properties of BL Lacertae described above?
   Let us start considering the radio-galaxies in the 3CR sample, see
   Fig. \ref{3c}. The extended radio luminosity of BL Lacertae is at the
   faint end of the FR~I in the 3CR, with only the nearby source 3C~272.1
   (M~84) being fainter.  From the point of view of the narrow emission lines,
   the [\ion{O}{III}] luminosity of BL Lacertae
   lies instead at the high end of 3CR/FR~I. Thus this source is a strong
   outlier (by a factor of $\sim$ 200--400 depending on the adopted value for
   its radio luminosity) from the relationship between line and radio
   luminosities followed by 3CR radio-galaxies\footnote{Other two strong
     outliers from the line-radio correlation are present in the 3CR, namely
     3C~371 and 3C~084; intriguingly, 3C~371 is a well-known BL Lac, while
     3C~084 (although often classified differently) hosts a highly polarized
     and variable optical nucleus \citep{martin76}, typical of BL Lac
     objects.}. This is true not only considering FR~I, but also from a
   comparison of the LEG and HEG classes, and casts substantial doubts on the
   association of BL Lacertae with FR~I, but more in general with the
   radio-galaxies in the 3CR sample. Note that also other classical samples of
   radio galaxies, like the B2 and the 2 Jy \citep{morganti97,tadhunter98}
   follow relations between radio and line emission similar to the 3CR.

   \citet{bal09} showed that a large ratio of line emission to the radio power
   with respect to 3CR sources is characteristic of the radio-loud AGN
   selected by \citet{best05a}. The latter authors cross-matched the $\sim
   212000$ galaxies drawn from the SDSS-DR2 with the NVSS and
   FIRST\footnote{Sloan Digital Sky Survey, Data Release 2 \citep{york00},
     National Radio Astronomy Observatory (NRAO) Very Large Array (VLA) Sky
     Survey \citep{condon98}, and the Faint Images of the Radio Sky at Twenty
     centimeters survey \citep{becker95}, respectively.} radio surveys,
   selecting a sample of 2215 radio-loud AGN (with a radio flux threshold of 5
   mJy) to which we refer hereafter as the SDSS/NVSS sample. \citet{bal10}
   explored the spectro-photometric properties of the SDSS/NVSS objects
   showing that they are generally hosted by massive early-type galaxies with
   a low excitation emission line spectrum. From the point of view of its
   narrow line luminosity and extended radio power BL Lacertae falls well
   within the region covered by the SDSS/NVSS sample (see Fig.\ \ref{3c}),
   although at a slightly higher $L \rm [\ion{O}{III}]$ than the region of
   higher galaxies density.
   
   However, these SDSS/NVSS sources do not show prominent broad lines like
   those observed in BL Lacertae. Because from the point of view of its radio
   flux ``misaligned BL Lacertae" objects would be included in the SDSS/NVSS
   catalogue even if located up to a redshift of $\sim 0.2$, the only
   possibility before we conclude that BL Lacertae is a unique object, at
   least in the nearby Universe, is that ``misaligned BL Lacertae" objects
   were rejected on an optical basis.

   Indeed, \citet{best05a} excluded from their sample sources recognised
   by the automated SDSS classification pipeline as QSO, because a bright nucleus prevents a detailed study of the host properties. These
   objects instead generally do show broad lines. Hence one may wonder
   whether it is possible that the parent population of BL Lacertae can be
   found among them. 
   We have then selected from the DR7 of the SDSS\footnote{Note that the DR7
     includes about four times more galaxies than the DR2 considered above.} all
   objects classified as QSO with redshift $z<0.1$, resulting in 1033 sources,
   90 of which have a radio source in the NVSS catalogue brighter than 5 mJy
   within 15$\arcsec$. We then looked for objects with properties similar to BL
   Lacertae, filtering out 42 sources with $r$ band and/or radio luminosities
   more than four times fainter than BL Lacertae, and 13 well-defined spiral
   galaxies, ending up with 35 candidates. 

   We downloaded their spectra from the SDSS public archive. Despite their
   automated classification as QSO, only 18 of them do show prominent broad
   emission lines\footnote{For the remaining 17 galaxies, we set upper limits
     to their broad \Ha\ luminosities fitting multiple emission lines to the
     spectra obtained after subtraction of the stellar component, as in
     \citet{but10}. All limits are between $\sim 10^{40}$ and $10^{41}$ erg
     s$^{-1}$.}. These galaxies, however, have [O~III] luminosity far higher
   than BL Lacertae, within the range $L \rm [\ion{O}{III}] \sim 2 \times
   10^{41} - 4 \times 10^{42}$ erg s$^{-1}$ (with a median of $\sim 5 \times
   10^{41}$ erg s$^{-1}$), a factor of 5 to 100 higher than in BL Lacertae,
   and are also characterized by a HEG spectrum.  These sources have line and
   radio luminosities as well as line ratios typical of Seyfert galaxies
   (e.g. \citealt{whittle85})\footnote{We remark one apparent
     exception, namely SDSS J122358.97+404409.3, with $L \rm [\ion{O}{III}]$
     only 20\% smaller than BL Lacertae, However, this object has a strong
     contamination by its active nucleus; the equivalent width (EW) of
     the NaD feature is EW(NaD) = 1.04 \AA, strongly reduced by the dilution
     of its nuclear emission with respect to a typical value for early type
     galaxies of $\sim 4$ \AA, indicating a nuclear contribution of $\sim$ 80
     \%. Its $r$ band magnitude corrected for the nuclear contribution is $M_R
     = -20.7, $ nearly 2 mag below the host of BL Lacertae.}.

Summarizing, we did not find any object in the SDSS/NVSS sample 
with isotropic properties similar to BL Lacertae. 
In particular, only a few sources show broad
emission lines and at the same time a
luminosity of the host galaxy and a radio power at least a quarter of BL
Lacertae; however, their narrow line luminosities exceed that of
BL Lacertae on average by a factor of 12, and their HEG spectrum contrasts
with the tentative identification of BL Lacertae as LEG.
Because, as mentioned above, ``misaligned BL Lacertae" objects would be found in
this extensive catalogue, we must conclude that we failed to find the parent
population of BL Lacertae, which apparently stands out as an ``orphan'' AGN.

This result contrasts with the requirement that to each object observed with
its jet (characterized by a bulk Lorentz factor $\Gamma$) forming a small
angle with our line of sight a parent population of
  mis-aligned objects must correspond. Although in presence of nuclear obscuration not all of
  them would show broad lines, this population is expected to be formed by $N
  \sim \Gamma^2 \sim 16 -1000 $ sources, having assumed a range of $\Gamma \sim
  4 - 30$ from \citet{jor05}. 

\section{Summary and conclusions}

More than a decade ago \citet{ver95} and subsequently \citet{cor96,cor00}
detected a broad H$\alpha$ emission line in the spectra of BL Lacertae. This
luminous line should have been detected in previous spectra, suggesting that
its flux must have increased by at least a factor 5 since 1989. The luminous
H$\alpha$ line suggested a Seyfert-like nucleus in BL
Lacertae, complicating the already difficult task of understanding what AGN
population this object (and BL Lac objects in general) belongs to, considering
that FR~I generally do not show broad lines.

To investigate this matter we acquired low- and high-resolution spectra of BL
Lacertae with the TNG during four nights in 2007--2008, when the source
optical brightness was $R \sim 14$--14.5.  Our spectra confirm the presence of
a luminous H$\alpha$ broad line of $\sim 4 \times 10^{41} \, \rm erg \,
s^{-1}$ and $\rm FWHM \sim 4600 \rm \, km \, s^{-1}$, as well as several
narrow emission lines. Through a critical comparison of our data with those
by \citet{cor00}, we concluded that the BLR luminosity has increased by $\sim
50\%$ in about ten years. This level of variability is not unusual for Broad
Lined AGN and it does not necessarily implies an evolutionary trend.

Then we examined the nuclear properties of BL Lacertae.  The relationship
between the SMBH mass and bulge luminosity in AGNs allowed us to derive a mass
of 4--$6 \times 10^8 \, M_{\sun}$.  Using the spectroscopic information to
calculate the virial mass, we instead obtained a value about 20--30 times
lower. To understand the reason of this discrepancy we analysed the disc
and BLR properties of other AGNs, and concluded that the BLR of BL Lacertae is
underluminous by a factor 70--300.

Finally, we analysed the physical quantities that do not depend on orientation
and beaming, and thus should also characterize the parent population of BL
Lacertae. We defined diagnostic indices with the most reliable narrow emission
lines, and found that their values provide a tentative identification of BL
Lacertae as a LEG. Broad lines are instead observed only in HEG, but the
diffuse radio luminosity of BL Lacertae is at least 100 times lower than in
these powerful radio sources. On the other hand miniature radio galaxies are
LEG, share both the narrow line and radio power properties of BL Lacertae, but
they do not show a BLR. Taking into account how the miniature radio galaxy
sample was selected, we expect that it should include ``misaligned BL
Lacertae" objects, unless they were excluded on the base of a QSO
appearance. An analysis of the galaxy morphology, spectral features, and radio
power of the QSO sources, initially discarded from the sample of miniature
radio-galaxies, revealed that no object meets the requirements to represent
the BL Lacertae parent population. Yet, for typical values of the Lorentz
factor, we would expect 10-10$^3$ ``misaligned BL Lacertae".

This leaves us with the only possibility that the observed properties of BL
Lacertae are the result of a transient short lasting phase. We can envisage
the following scenario, somewhat similar to that already suggested by
\citet{cor96}. BL Lacertae in its initial state has properties similar to the
sources of the SDSS/NVSS sample. Indeed, these are massive early-type galaxies
and a large number of them have narrow lines and radio luminosities similar to
that of BL Lacertae. From the point of view of their optical spectra they are
LEG and lack broad lines.  Subsequently (possibly $\sim 20$ years ago), its
BLR underwent an increase of luminosity due to an increased amount of cold
gas in the nuclear regions and/or to a higher level of ionizing
continuum. These two effects may even be related and caused by a fresh input
of accreting gas. The BLR structure might not have yet reached a stable
configuration, accounting for its different properties when compared to other
AGN.  Also the NLR luminosity will grow with time and will also eventually change
its state of ionization, but on a much larger timescale with respect to the
BLR.

Based on the analysis of a single object it is clearly impossible to set a
timescale for the duration of the putative bright phase. Furthermore, BL
Lacertae was probably discovered since this object has been subject to
repeated spectroscopic observations. However, our failure to find objects in
the local Universe that might constitute its parent population suggests that
the timescale associated with the period of high accretion must be orders of
magnitude shorter than the lifetime of radio-loud AGN.

An alternative possibility is that the birth of the BLR marks the transition
from a low-power radio galaxy to a high-power source. This would require a
rapid increase in the luminosity of the large-scale radio structures to
reach the level observed in e.g.\ the HEG of the 3CR sample, within a
sufficiently short time so as not to produce a substantial population of
transient sources.  Instead, the available data rule out that BL Lacertae
became an AGN only very recently, i.e.\ that we are witnessing its birth,
because its radio emission extends $\sim 10$ kpc away from the core. This
implies that this source is active since at least $\sim 3 \times 10^5$ years,
assuming an expansion speed of 0.1 c.

We conclude that the parent population of BL Lacertae can be found among the
large population of miniature radio-loud AGN forming the SDSS/NVSS sample, but
this also requires that this object is experiencing a short transient
phase. A continuation of the spectroscopic monitoring of this peculiar source
caught in a crucial phase of its evolution can help us tremendously in our
study of the physics and evolution of these systems.

\begin{acknowledgements}
  We thank David J. Axon and the anonymous referee for their useful comments
  and suggestions. Partly based on data taken and
  assembled by the WEBT collaboration and stored in the WEBT archive at the
  Osservatorio Astronomico di Torino - INAF
  (http://www.oato.inaf.it/blazars/webt/).
  This research has made use of the NASA/IPAC
  Extragalactic Database (NED) which is operated by the Jet Propulsion
  Laboratory. California institute of Technology, under contract with the
  National Aeronautics and Space Administration.  This research has made use
  of NASA's Astrophysics Data System (ADS). This research has made use of the
  SDSS Archive, funding for the creation and distribution of which was
  provided by the Alfred P. Sloan Foundation, the Participating Institutions,
  the National Aeronautics and Space Administration, the National Science
  Foundation, the U.S. Department of Energy, the Japanese Monbukagakusho, the
  Max Planck Society, and The Higher Education Funding Council for England.
\end{acknowledgements}

\end{document}